\documentstyle[amsfonts,12pt,thmsa,sw20bams]{article}

\input tcilatex

\begin{document}

\author{Choulakian, V. \and Universit\'{e} de Moncton, Moncton, N.B., E1A
3E9, Canada. \and email: vartan.choulakian@umoncton.ca}
\title{Some Numerical Results on the Rank of Generic Three-Way Arrays over $%
{\bf 
\mathbb{R}
}$}
\date{1 June 2009}
\maketitle

\begin{abstract}
The aim of this paper is the introduction of a new method for the numerical
computation of the rank of a three-way array, ${\bf X}\in {\bf 
\mathbb{R}
}^{I\times J\times K},$ over ${\bf 
\mathbb{R}
}$. We show that the rank of a three-way array over ${\bf 
\mathbb{R}
}$ is intimately related to the real solution set of a system of polynomial
equations. Using this, we present some numerical results based on the
computation of Gr\"{o}bner bases.


\end{abstract}

\section{Introduction}

Let ${\bf X}\in {\bf 
\mathbb{R}
}^{I\times J\times K}$ be a tensor of order 3, sometimes named a three-way
array or a three-mode data set. A rank 1 or a decomposed tensor is 
\begin{equation}
{\bf D}={\bf a}\otimes {\bf b}\otimes {\bf c},  \tag{1}
\end{equation}%
where ${\bf a}\in {\bf 
\mathbb{R}
}^{I},\ {\bf b}\in {\bf 
\mathbb{R}
}^{J}$ and ${\bf c}\in {\bf 
\mathbb{R}
}^{K}$, and $\otimes $ is the tensor product, sometimes named also outer
product$.$ ${\bf X}${\bf \ }can be expressed as a sum of decomposed tensors
given in (1),%
\begin{equation}
{\bf X}=\sum_{\alpha =1}^{r}{\bf D}_{\alpha }.  \tag{2}
\end{equation}%
The rank of ${\bf X}$ is defined to be the minimal integer $r,$ see for
instance Kruskal (1977, 1989). In data analysis, this implies that the rank
of a three-way array is the smallest number of components that provide a
perfect fit in Candecomp/Parafac (CP), see for instance, (Carroll and Chang,
1970, and Harshman, 1970). In statistics CP is considered a natural
extension of singular value decomposition or principal components analysis
to three-way data.

There is quite a literature concerning the value of maximal rank, generic
rank or typical rank of three-way arrays in the area of statistics,
algebraic complexity theory and algebraic geometry. Some references, among
others, are: Ja' Ja' (1979), Kruskal (1977, 1983, 1989), Strassen (1983),
Ten Berge (1991, 2000, 2004a, 2004b), Ten Berge and Kiers (1999), Ten Berge
and Stegeman (2006), Comon and Ten Berge (2008), B\"{u}rgisser et al (1997),
Catalisano et al (2002), Friedland (2008) and Abo et al. (2006). Friedland
(2008) provides an up todate survey with some new results on the generic
rank of three-way arrays.

First, we give the following

{\bf Definition 1}: A dataset is called {\it generic} if its elements are
randomly generated from a continuous distribution.

The generic and typical ranks are defined in the following way by Comon and
Ten Berge (2008): Given that the rank of $I\times J\times K$ arrays is
bounded, one can partition the arrays according to the rank values. Generic
rank is defined to be true almost everywhere; while typical ranks are
associated with the rank values that occur with positive probability. So, if
there is a single typical rank, then it may be called generic rank; that is,
a generic rank is typical, but the converse is not true.

Ten Berge (2000) classified three-way arrays into three classes: very tall,
tall and compact. Let ${\bf X}\in {\bf 
\mathbb{R}
}^{I\times J\times K}$ be a tensor of order 3 with $I\geq J\geq K$. The array%
$\ {\bf X}$ is called very tall when $I\geq KJ;$ ${\bf X}$ is tall when $%
KJ-J<I\leq KJ-1;$ ${\bf X}$ is compact when $I\leq KJ-J.$ The generic rank
of the very tall arrays is very well known and easiest to prove: it is $KJ$.
Ten Berge (2000) showed that all tall three-way arrays have generic rank $I$%
; and the tallest among the compact arrays, that is when $I=KJ-J,$ have
typical rank $\left\{ I,I+1\right\} .$\ Ten Berge and Stegeman (2006)
provided some further results on the compact case. Friedland (2008) showed
that: typical rank($12\times 4\times 4{\bf )\geq }12,$ typical rank($%
11\times 4\times 4{\bf )\geq }11,$ and typical rank($I\times J\times K)\geq
I\ $for\ $(I,J,K)=((J-1)^{2}+1,J,J)\ $when\ $J\geq 2$. These results are all
based on mathematical proofs. However, the rank computation problem has also
been approached from a numerical point of view: Comon and ten Berge (2008)
and Friedland (2008) applied Terracini's lemma, based on the numerical
calculation of the maximal rank of the Jacobian matrix of (2), to obtain
numerically the generic rank of some three-way arrays. The numerical method
based on Terracini's lemma, when used to evaluate rank over ${\bf 
\mathbb{R}
,}$ gives the generic rank when the typical rank is single-valued, and the
smallest typical rank value otherwise.

Two well known facts are: a) There is no known method to calculate the rank
of a given three-way dataset, Martin (2004, AIM tensor workshop); b) A
three-way array over ${\bf 
\mathbb{R}
}$ may have a different rank than the same array considered over $%
\mathbb{C}
,$ (Kruskal, 1989).

We shall be concerned by the numerical computation of the rank of a
three-way array over ${\bf 
\mathbb{R}
}$ only.

Computationally, the most primitive approach to the numerical evaluation of
the typical rank of three-way arrays is based on the alternating least
square (ALS) minimization algorithm: It is to run ALS many times to
convergence on many generic three-way arrays of a given format, and to check
whether or not the fit is perfect for a given number of components. But as a
referee remarked, this approach has 2 problems: First, we do not know how
many three-way arrays of a given format to examine before a valid inference
can be drawn. For instance, when 100000 arrays have been examined and all
seem to have the same rank $\alpha $, it does not follow that $\alpha $ is
indeed the generic rank for that array format. After all, a different rank
may occur with an extremely small yet positive probability.\ Second, the
decision of when to terminate ALS is hazardous, because even if the residual
sum of squares is, say, $exp(-32)$, this does not prove that it is zero; in
fact, it may have zero as infimum. The present paper relieves us from both
above mentioned problems: It offers a straightforward method of determining
the rank of any given array over ${\bf 
\mathbb{R}
}$, based on inspection of the number of real roots of a system of certain
polynomial equations.

The real solution set of a system of polynomial equations is called
semi-algebraic set in real algebraic geometry, see Basu, Pollack and Roy
(2006) or Friedland (2008). Semi-algebraic sets are open sets and are
composed of a finite union of connected components, where each component is
called a basic semi-algebraic set. The main problem can be reformulated as:
For a given tensor ${\bf X}$ over ${\bf 
\mathbb{R}
}$ calculate the number of connected components where each component is
characterized by a unique real rank value. Our numerical results will shed
some light on this. The numerical results on simulated datasets will be
obtained by computing the Gr\"{o}bner bases using Maple 12 of the system of
polynomial equations characterizing the dataset. We note that generic
datasets and random numbers are generated from integers between $-99$ and $%
99 $.

The paper is organized as follows. In section 2 we present the main lemma
which provides a necessary and sufficient condition that a three-way array
can be expressed as a sum of a fixed number of decomposed tensors. All
results in this paper will be based on this lemma. In section 3, we show how
the lemma can be applied to compute the rank of a generic tensor over ${\bf 
\mathbb{R}
}$ numerically for some cases. In section 4, we show another application of
the lemma for the computation of rank for nongeneric particular datasets. In
section 5, we show how the lemma can be applied to compute the rank of
generic $I$ symmetric $J\times J$ arrays, named INDSCAL arrays, over ${\bf 
\mathbb{R}
.}$ And finally in section 6 we conclude.

\section{Main Lemma}

Let ${\bf X}\in {\bf 
\mathbb{R}
}^{I\times J\times K}$ be a three-way dataset and $2\leq K\leq J\leq I$. The
lemma provides a necessary and sufficent condition that the tensor ${\bf X}$
can be expressed as a sum of $I$ decomposed tensors; that is 
\begin{eqnarray}
{\bf X} &=&\sum_{\alpha =1}^{I}{\bf D}_{\alpha },  \nonumber \\
&=&\sum_{\alpha =1}^{I}{\bf a}_{\alpha }\otimes {\bf b}_{\alpha }\otimes 
{\bf c}_{\alpha },  \TCItag{3}
\end{eqnarray}%
where $\{{\bf a}_{\alpha }|$ $\alpha =1,...,I\}$ is a basis for ${\bf 
\mathbb{R}
}^{I},$ ${\bf c}_{\alpha }\in {\bf 
\mathbb{R}
}^{K}$ and$\ {\bf b}_{\alpha }\in {\bf 
\mathbb{R}
}^{J}$. Note that if (3) is true, then $rank({\bf X)\leq }I$. We denote by $%
{\bf X}_{k}\in {\bf 
\mathbb{R}
}^{I\times J}$ the $k$th slice in ${\bf X}$ for $k=1,...,K.$We note that (3)
can be written as%
\begin{eqnarray}
{\bf X}_{k} &=&\sum_{\alpha =1}^{I}c_{k\alpha }{\bf a}_{\alpha }\otimes {\bf %
b}_{\alpha }\text{ \ \ for \ \ }k=1,...,K,  \nonumber \\
&=&{\bf AD(c}_{k}{\bf )B}^{\prime }\text{ \ \ for \ \ }k=1,...,K,  \TCItag{4}
\end{eqnarray}%
where ${\bf A}=({\bf a}_{1}\ {\bf a}_{2}...{\bf a}_{I})\in {\bf 
\mathbb{R}
}^{I\times I},$ ${\bf B}=({\bf b}_{1}\ {\bf b}_{2}...{\bf b}_{I})\in {\bf 
\mathbb{R}
}^{J\times I}$, ${\bf C}=(c_{k\alpha })\in {\bf 
\mathbb{R}
}^{K\times I}$ and ${\bf D(c}_{k}{\bf )}={\bf Diag}({\bf c}_{k})\in {\bf 
\mathbb{R}
}^{I\times I}$ is a diagonal matrix with diagonal elements $c_{k\alpha }.$
Note that the vector ${\bf c}_{k}\in {\bf 
\mathbb{R}
}^{I}$ represents the $k$th row of ${\bf C.}$

We consider the system of polynomial equations%
\begin{equation}
{\bf s}_{\alpha }^{\prime }{\bf X}_{k}=c_{k\alpha }{\bf b}_{\alpha }^{\prime
}\text{ \ \ for \ }k=1,...,K\text{ and }\alpha =1,...,I,  \tag{5}
\end{equation}%
where $\{{\bf s}_{\alpha }|$ $\alpha =1,...,I\}$ is a basis for ${\bf 
\mathbb{R}
}^{I}$ and ${\bf c}_{\alpha }\in {\bf 
\mathbb{R}
}^{K},\ $and ${\bf b}_{\alpha }\in {\bf 
\mathbb{R}
}^{J}$. We note that (5) can be written as 
\begin{equation}
{\bf S}^{\prime }{\bf X}_{k}={\bf D(c}_{k}{\bf )B}^{\prime }\text{ for }%
k=1,...,K,  \tag{6}
\end{equation}%
where ${\bf S}$ has columns ${\bf s}_{\alpha }.\bigskip $

{\bf Lemma 1}: (6) is a necessary and sufficient condition for (4).

{\em Proof}: Let ${\bf I}={\bf AS}^{\prime }$, then (4) is true if and only
if (6) is true.\bigskip

{\bf Remark 1}: a) To see if (5) is true, we solve the system of polynomial
equations%
\begin{equation}
{\bf s}^{\prime }{\bf X}_{k}=c_{k}{\bf b}^{\prime }\text{ \ \ for \ }%
k=1,...,K,  \tag{7}
\end{equation}%
for ${\bf s}\in {\bf 
\mathbb{R}
}^{I},\ {\bf b}\in {\bf 
\mathbb{R}
}^{J}$ and ${\bf c}\in {\bf 
\mathbb{R}
}^{K}.$

b) We note that (7) has two indeterminacies: It can be rewritten as ${\bf s}%
_{\ast }^{\prime }{\bf X}_{k}=c_{k\ast }{\bf b}_{\ast }^{\prime }$ \ \ for \ 
$k=1,...,K,$ , where for instance, ${\bf s}_{\ast }=\lambda {\bf s}$ for any
scalar $\lambda \neq 0$ $,\ c_{k\ast }=\mu c_{k}$ for any scalar $\mu \neq
0, $ and ${\bf b}_{\ast }=\lambda {\bf b/}\mu {\bf .}$ To eliminate these
indeterminacies, hereafter, we fix 
\begin{equation}
c_{1}=1\text{ \ \ and \ \ }s_{I}=1.  \tag{8}
\end{equation}

c) Theorem 2.4 of Friedland (2008) provides another characterization for (3)
or (4): It states that each slice ${\bf X}_{k}\in span({\bf a}_{1}\otimes 
{\bf b}_{1},...,{\bf a}_{I}\otimes {\bf b}_{I})$ and the rank$({\bf X)}$
equals the minimal dimension of the $span({\bf a}_{1}\otimes {\bf b}_{1},...,%
{\bf a}_{I}\otimes {\bf b}_{I}).$

d) The necessary condition, when rank$({\bf X)=}I{\bf ,}$ which was shown to
be also sufficient afterwards, was used many times by Ten Berge and his
coworkers, Ten Berge (2000), Ten Berge (2004a), and Ten Berge, Sidiropoulos
and Rocci (2004).

\section{Rank computation}

We shall suppose in the sequel that ${\bf X}\in {\bf 
\mathbb{R}
}^{I\times J\times K}$ is a generic three-way array and $2\leq K\leq J\leq
I\leq KJ$. Then we have the following well known inequality: rank$({\bf %
X)\geq }I.$ We will check if ${\bf X}$ has rank $I.$\ By the Main Lemma ,
the tensor ${\bf X}$ has rank $I$, if for parameter vectors ${\bf s}\in {\bf 
\mathbb{R}
}^{I},\ {\bf b}\in {\bf 
\mathbb{R}
}^{J}$ and ${\bf c}\in {\bf 
\mathbb{R}
}^{K}$ the system of polynomial equations (7) subject to (8) have $I$ real
solutions $(c_{k\alpha }{\bf ,}\ {\bf b}_{\alpha },$ ${\bf s}_{\alpha })$
for $\alpha =1,...,I$, such that the elements of the set $\{{\bf s}_{\alpha
}|$ $\alpha =1,...,I\}$ is a basis for ${\bf 
\mathbb{R}
}^{I};$ that is, (7) with (8) has $I$ real isolated solutions. Let us see
how can we know if this is true. The system of polynomial equations (7) with
(8) is equivalent to

\begin{equation}
{\bf s}^{\prime }({\bf X}_{k}-c_{k}{\bf X}_{1})={\bf 0}^{\prime }\text{\ for
\ }k=2,...,K.  \tag{9}
\end{equation}%
So the number of equations, $neq$, in (9) is%
\begin{equation}
neq=(K-1)J,  \tag{10}
\end{equation}%
and the number of degrees of freedom or the number of free variables , $df,$
is%
\begin{equation}
df=(I-1)+(K-1),  \tag{11}
\end{equation}%
because of (8) there are $(K-1)$ free $c_{k}$ 's and $(I-1)$ free $s_{i}$ 's.

We are interested in the study of the number of solutions of (9) over ${\bf 
\mathbb{R}
}$ for {\it generic} data. We distinguish three cases named, minimal when $%
neq=df$, overdetermined when $df<neq,$ and, underdetermined when $df>neq$.
We note that Abo et al. (2006) also distinguished three cases that they
named subabundant, superabundant and equiabundant: these were used for
induction purposes.

\subsection{{\bf Case 1: Minimal System(}${\bf neq=df)}$}

When $I=(K-1)(J-1)+1$, $neq=df$, and the system (9) is called minimal. The
number of real solutions is {\it bounded}; an upper bound is provided by
Khovanskii's theorem, see Sturmfels (2002),\bigskip

{\bf Theorem 1 (Khovanskii): } Consider $n$ polynomials in $n$ variables
involving $m$ distinct monomials in total. The number of isolated roots in
the positive orthant ($R_{+})^{n}$ of any such system is at most $%
2^{(_{2}^{m})}(n+1)^{m}.\bigskip $

In our case $n=neq=df=(K-1)J$ and $m=I-1+(K-1)(I-1)=K(I-1).$ The number of
isolated roots in the positive orthant ($R_{+})^{df}$ of any such system is
at most $2^{(_{2}^{m})}(df+1)^{m}$ and $\ m$ is the number of distinct
monomials in the system (9). So (9) may or may not have $I$ real isolated
solutions. In case (9) has $I$ real isolated solutions, then rank$({\bf X}%
)=I $; otherwise we embed it, which is discussed later on.

{\bf Example 1: }$I\times I\times 2${\bf \ arrays: }$neq=df=I$

This class of arrays is discussed in detail by Ten Berge (1991), who showed
that the typical rank of such arrays is $\{I,I+1\}$. To check if the rank of
a generic $I\times I\times 2$ array is $I,$ it suffices to solve (9), which
reduces to finding the real roots of the determinantal equation $det({\bf X}%
_{2}-c_{2}{\bf X}_{1})=0.$ If $det({\bf X}_{2}-c_{2}{\bf X}_{1})=0$ has $I$
real roots, then $rank({\bf X})=I,$ otherwise $rank({\bf X})=I+1.$
Simulation results for $5000$ generic $3\times 3\times 2$ arrays produced
one real root $51.76\%$ and $3$ real roots $48.24\%$ of the time. So we can
deduce that Pr(rank ($3\times 3\times 2${\bf \ }array$)=3)\approx 48.24\%$
and Pr(rank ($3\times 3\times 2${\bf \ }array$)=4)\approx 51.76\%.\medskip $

{\bf Example 2: }$I\times J\times 3${\bf \ arrays with }$I=2J-1${\bf : }$%
neq=df=2J$

a) $5\times 3\times 3${\bf \ arrays: }$neq=df=6.$ This class of arrays is
also discussed in Ten Berge (2004a), where Ten Berge showed that generic 
{\bf \ }$5\times 3\times 3${\bf \ }arrays have either rank 5 or rank 6 with
positive probability. Further, he showed that a closed form solution for the
case when the array has rank 5 corresponds to finding the number of real
roots of a sixth degree polynomial equation: if there are 6 real roots, then
the array has rank 5, otherwise its rank is 6. Table 1 displays the number
of real roots obtained by solving the system (9) for $1000$ simulated
generic arrays. First, we note that the solution set of (9) always admitted
6 roots, as expected according to Ten Berge (2004a); further, the number of
real solutions is an even number or zero. Second, Pr(rank ($5\times 3\times
3 ${\bf \ }array$)=5)\approx 6.8\%$ and Pr(rank ($5\times 3\times 3${\bf \ }%
array$)=6)\approx 93.2\%.\medskip $

\begin{tabular}{|l|l|l|l|l|}
\hline
\multicolumn{5}{|l|}{{\bf Table 1: Simulation results for 1000 generic }$%
5\times 3\times 3${\bf \ arrays}.} \\ \hline
$\func{real}$ $roots$ & 0 & 2 & 4 & 6 \\ \hline
$counts$ & 47 & 501 & 384 & 68 \\ \hline
\end{tabular}%
\medskip\ 

b) $7\times 4\times 3${\bf \ arrays: }$neq=df=8.$ Table 2 displays the
number of real roots obtained by solving the system (9) for $1000$ simulated
generic arrays. First, we note that the solution set of (9) always admitted
10 roots and the number of real solutions is an even number or zero. Second,
Pr(rank ($7\times 4\times 3${\bf \ }array$)=7)\approx 4.2\%$.$\medskip $

$%
\begin{tabular}{|l|l|l|l|l|l|l|}
\hline
\multicolumn{7}{|l|}{{\bf Table 2: Simulation results for 1000 generic }$%
7\times 4\times 3${\bf \ arrays}.} \\ \hline
$\func{real}$ $roots$ & 0 & 2 & 4 & 6 & 8 & 10 \\ \hline
$counts$ & 16 & 268 & 456 & 218 & 40 & 2 \\ \hline
\end{tabular}%
\medskip $

c) $9\times 5\times 3${\bf \ arrays: }$neq=df=10.$ Table 3 displays the
number of real roots obtained by solving the system (9) for $1000$ simulated
generic arrays. First, we note that the solution set of (9) always admitted
15 roots and the number of real solutions is an odd number. Second, Pr(rank (%
$9\times 5\times 3${\bf \ }array$)=9)\approx 6\%$ and Pr(rank ($9\times
5\times 3${\bf \ }array$)=10)\approx 94\%$. This latter result follows from
Ten Berge (2000, Result 5) or see Example 5 describing tallest compact
arrays, by embedding $9\times 5\times 3${\bf \ }arrays into $10\times
5\times 3${\bf \ }arrays.

$%
\begin{tabular}{|l|l|l|l|l|l|l|l|}
\hline
\multicolumn{8}{|l|}{{\bf Table 3: Simulation results for 1000 generic }$%
9\times 5\times 3${\bf \ arrays}.} \\ \hline
$\func{real}$ $roots$ & 1 & 3 & 5 & 7 & 9 & 11 & 13 \\ \hline
$counts$ & 34 & 290 & 404 & 212 & 51 & 8 & 1 \\ \hline
\end{tabular}%
$\medskip\ 

{\bf Example 3: }$I\times J\times 4${\bf \ arrays with }$I=3J-2${\bf : }$%
neq=df=3J$

Numerical computations showed that $\neq (roots$ of $10\times 4\times 4$
arrays$)=20;$ $\neq (roots$ of $13\times 5\times 4$ arrays$)=35$ and $\neq
(roots$ of $16\times 6\times 4$ arrays$)=56.$ Table 4 shows that Pr(rank ($%
10\times 4\times 4${\bf \ }array$)=10)\approx 7.8\%$.\medskip

$%
\begin{tabular}{|l|l|l|l|l|l|l|l|l|}
\hline
\multicolumn{9}{|l|}{{\bf Table 4: Simulation results for 1000 generic }$%
10\times 4\times 4${\bf \ arrays}.} \\ \hline
$\func{real}$ $roots$ & 0 & 2 & 4 & 6 & 8 & 10 & 12 & 14 \\ \hline
$counts$ & 2 & 78 & 284 & 342 & 216 & 58 & 14 & 6 \\ \hline
\end{tabular}%
\medskip $

{\bf Remark 2}: a) To calculate a Gr\"{o}bner basis for (9) in Example 2 for%
{\bf \ }$I\times J\times 3${\bf \ }arrays with{\bf \ }$I=2J-1$, we used pure
lexicographic order given by the following sequence ($%
s_{1},...,s_{I-1},c_{3},c_{2})$ of the free variables. In all cases the Gr%
\"{o}bner basis, denoted by $G_{\beta },$ consisted of $(K-1)J$ polynomials
having the following form: $G_{1}(c_{2})=0$, $G_{2}(c_{2},c_{3})=$ poly$%
_{2}(c_{2})+c_{3}=0,$ $G_{3,\alpha }(c_{2},s_{\alpha })=$ poly$_{\alpha
}(c_{2})+s_{\alpha }=0$ for $\alpha =1,...,I-1.$ It is important to note
that this particular form of the Gr\"{o}bner basis polynomials, $G_{\beta },$
shows that the degree of the polynomial $G_{1}(c_{2})=0,$ denoted by $%
degG_{1}(c_{2}),$ represents the number of roots of the system (9). An
introduction to Gr\"{o}bner basis can be found in, among others, Cox et al.
(2007). Example 6 show quite in detail the Gr\"{o}bner basis application to
a generic array.

b) The Maple 12 commands to do the computations in Example 2 are shown in
Appendix 1.

c) For $I\times I\times 2$ arrays and $I\geq 2,\ $ $det({\bf X}_{2}-c_{2}%
{\bf X}_{1})=G_{1}(c_{2})=0,$ where $G_{1}(c_{2})=0$ is the first element of
the Gr\"{o}bner basis. This phenomenon will be also seen for tallest compact
arrays, see Examples 4, 5 and 6.\medskip

A reviewer noted that the right hand side of (7) is a Segre variety, which
is the image of the Segre map, $\Sigma _{(K-1),(J-1)}$. The Segre map sends
an element of the projective space $P^{(K-1)}\times P^{(J-1)}$ into $%
P^{KJ-1}.$ While the left hand side of (7) is a linear space of projective
dimension $I-1=(K-1)(J-1).$ So, (7), will represent the intersection of the
linear space with the Segre map, and the number of intersections is the
degree of the Segre variety given in (12), see for instance Harris (1992, p.
233). This result is summarized in the following \bigskip

{\bf Theorem 2:} Let $I=(K-1)(J-1)+1$ and $2\leq K\leq J\leq I,$ then for
generic data the number of roots (real or complex) of the polynomial system
(9) is%
\begin{equation}
degG_{1}(c_{2})=\binom{K-1+J-1}{K-1}.  \tag{12}
\end{equation}%
\bigskip

{\bf Corollary 1}: For minimal systems and $3\leq K\leq J\leq I$, $%
I<degG_{1}(c_{2}).$

{\em Proof:} Let $n=J-1$ and $m=K-1$. We have to show that%
\[
mn+1\leq \frac{(m+n)!}{n!m!}\text{ \ \ \ for \ \ }2\leq m\leq n. 
\]
It is true for $m=2$. For $m\geq 3,$ we have%
\begin{eqnarray*}
\frac{(m+n)!}{n!m!} &=&\left[ \frac{(n+1)}{m}\frac{(n+2)}{m-1}...\frac{%
(n+m-2)}{3}\right] \left[ \frac{(n+m-1)}{2}\frac{(n+m)}{1}\right] \\
&\geq &\left[ \frac{(n+m-1)}{2}\frac{(n+m)}{1}\right] .
\end{eqnarray*}%
So, it is sufficient to show that $(n+m)(n+m-1)\geq 2(mn+1),$ which is
easily seen to be true.\bigskip

{\bf Corollary 2}: The typical rank of arrays with a minimal system have
more than one rank value and the minimum attained value is $I$.

{\em Proof: }The rank of a generic array with a minimal system is $I$, if
the number of real roots of $G_{1}(c_{2})$ is greater than or equal to $I$;
otherwise its rank is greater than $I$.\bigskip

We note that Corollary 2 generalizes Friedland (2008), who showed that:
typical rank($I\times J\times K)\geq I\ $for\ $(I,J,K)=((J-1)^{2}+1,J,J)\ $%
when\ $J\geq 2$.

\subsection{{\bf Case 2: Underdetermined System(}${\bf df>neq})$}

When\ $(K-1)(J-1)+2\leq I\leq IJ,$ $df>neq,$ and the system (9) is called
underdetermined. The upper bound for the number of isolated roots of (9) is%
{\it \ infinity; so }(9) may or may not have $I$ real isolated solutions: So
the attained minimum bound for the rank of a generic three-way array is, $%
b_{\min }=I.$ Before discussing two general classes studied in detail by Ten
Berge (2000), we introduce some notation.

The system (9) can be written as%
\begin{equation}
{\bf s}^{\prime }{\bf \Gamma }={\bf s}^{\prime }\left[ ({\bf X}_{2}-c_{2}%
{\bf X}_{1}),({\bf X}_{3}-c_{3}{\bf X}_{1}),...,{\bf X}_{K}-c_{K}{\bf X}_{1}%
\right] {\bf =0}^{\prime }{\bf ,}  \tag{13}
\end{equation}%
where the number of columns of the matrix ${\bf \Gamma }$ is 
\begin{eqnarray}
n\func{col}{\bf \Gamma } &{\bf =}&{\bf (}K-1{\bf )}J,  \TCItag{14} \\
&=&neq;  \nonumber
\end{eqnarray}%
and the number of rows of ${\bf \Gamma }$ is 
\begin{equation}
n\func{row}{\bf \Gamma }{\bf =}I,  \tag{15}
\end{equation}%
and ${\bf \Gamma }$ is a matrix function of the parameters $c_{2},...,c_{K}.$
We also define 
\begin{equation}
nbil=K-1,  \tag{16}
\end{equation}%
which represents the minimal number of $c_{k}$\ parameters that can be
specialized to make the system of polynomial equations (13) linear. In
algebraic geometry, the replacement of variables by specific values is
called specialization.

{\bf Example 4}: {\bf Tall arrays: }$df-neq\geq nbil$

These are arrays when $(K-1)J<I\leq KJ$ and $I\geq J\geq K,$ whose generic
rank is $I,$ as shown by Ten Berge (2000, Result 2). This implies that $%
(15)>(14)$, that is $I>{\bf (}K-1{\bf )}J{\bf ,}$ or, $df-neq\geq nbil=K-1$,
where $nbil$ is given in (16). By assigning random values to the $(K-1)c_{k}$%
's in (13), we reduce (13) to a system of linear equations, which will have
a solution for any generic data; so (13) will admit $I$ real and isolated
solutions; from which we deduce that the generic rank of tall arrays is $I$.

{\bf Example 5}: {\bf Tallest compact arrays: }$n\func{col}{\bf \Gamma =}n%
\func{row}{\bf \Gamma }$ and $K\geq 3$

These are arrays when $I=J(K-1),$ $\ I\geq J\geq K$ and $K\geq 3.$ Note that
we exclude $I\times I\times 2$ arrays for $I\geq 2$ discussed in Example 1$.$
Ten Berge (2000, Results 3, 4 and 5) discussed this case.

When $I=(K-1)J$ and $K\geq 3,$ it implies that $(14)=(15),$ that \ is, ${\bf %
\Gamma }$ is a square matrix. Solving (13) for $c_{k}$ 's \ for \ $k=2,...,K$
is equivalent to solving $det({\bf \Gamma )=}0.$ The leading monomial in $%
det({\bf \Gamma )=}0$ is $\dprod\limits_{k=2}^{K}c_{k}^{J}.$ If $J$ is an
odd integer, then (13) will have infinite number of real solutions: Assign
random continuous numbers to $c_{k}$ 's \ for \ $k=3,...,K,$ and solve for $%
c_{2}.$ This corresponds to Result 5 in Ten Berge (2000), which states: When 
$I=J(K-1)\ $and $\ I\geq J\geq K$ and $K\geq 3$ and $J$ is odd, then the
typical rank is $I$. If $J$ is an even integer, then (13) may have infinite
number of real solutions or finite number of real solutions or 0 real
solution: For instance for $J=4$ and $K=3$, the polynomial $%
f(c_{2},c_{3})=3c_{2}^{4}c_{3}^{4}+1$ has 0 real solution, the polynomial $%
f(c_{2},c_{3})=3c_{2}^{4}c_{3}^{4}-1$ has infinite number of real and
distinct solutions, and the polynomial $%
f(c_{2},c_{3})=3c_{2}^{4}(c_{3}^{4}-1)$ has a finite number of real
solutions. Ten Berge (2000) specifically discussed the case of $8\times
4\times 3$ arrays, where he stated that typical rank of such arrays is $%
\left\{ 8,9\right\} $ and for randomly sampled data the rank of 9 is
extremely rare. Similarly, Friedland(2008, Th.7.2) showed that typical rank
of $12\times 4\times 4$ arrays has more than one value. We conducted a
limited simulation study on generic $8\times 4\times 3$ and $12\times
4\times 4$ arrays; and each time we got $I$ real isolated solutions. The
simulation study was done in the following way: For a generic dataset let $%
f(c_{2},c_{3}...,c_{K})=det({\bf \Gamma )=}0;$ assign random values to the
parameters $c_{3}...,c_{K},$ then solve for $c_{2}.$ This shows that for
generic data, when $I=J(K-1)$ and $K\geq 3$ the rank is $I$ with very high
probability. Also, see example 6.

\subsection{Example 6}

We consider a simulated generic dataset of size $7\times 4\times 3$ having
the following three slices

{\bf X}$_{1}^{\prime }$ := $\left( 
\begin{array}{c}
\lbrack -50,-38,-98,-93,-32,8,44] \\ 
\lbrack -22,-18,-77,-76,-74,69,92] \\ 
\lbrack 45,87,57,-72,-4,99,-31] \\ 
\lbrack -81,33,27,-2,27,29,67]%
\end{array}%
\right) $

{\bf X}$_{2}^{\prime }$ := $\left( 
\begin{array}{c}
\lbrack 99,-25,24,-61,31,25,50] \\ 
\lbrack 60,51,65,-48,-50,94,10] \\ 
\lbrack -95,76,86,77,-80,12,-16] \\ 
\lbrack -20,-44,20,9,43,-2,-9]%
\end{array}%
\right) $

{\bf X}$_{3}^{\prime }$ := $\left( 
\begin{array}{c}
\lbrack 90,-82,29,52,42,-62,22] \\ 
\lbrack 80,-70,70,-13,18,-33,14] \\ 
\lbrack 19,41,-32,82,-59,-68,16] \\ 
\lbrack 88,91,-1,72,12,-67,9]%
\end{array}%
\right) $\bigskip

Our aim is to find the rank of {\bf X}, by representing it as in (6). This
dataset has a minimal system of polynomial equations. We solve equation (9)
via Gr\"{o}bner basis using the lexicograhic order ($%
s_{1},s_{2},s_{3},s_{4},s_{5},s_{6},c_{3},c_{2}).$ The first two polynomials
of the Gr\"{o}bner basis are

$G_{1}(c_{2})=0=$

$-$258797975083999058663603818114838724165583573114256294

$-$1987946767932180125365724555441379125561037244553323732*$c_{2}$

$-$7447583055793225423658520296174635567495579387052082486*$c_{2}^{2}$

+18477292423934054741969006645285810935999768448664319668*$c_{2}^{3}$

+162868576676248184458245504688648649537661407605447407344*$c_{2}^{4}$

+22324671325209561198922665813216562379249229294549244662*$c_{2}^{5}$

$-$93044594774454916354246852601811640731920664188422515202*$c_{2}^{6}$

+1034990365268175640254342731156079689724071145674294956746*$c_{2}^{7}$

$-$1399215109838269848671482913176200716552825195700591390075*$c_{2}^{8}$

+155346700794650490501115016130585172314320583287574900147*$c_{2}^{9}$

+645072630378953757678717001000719217821315452777261680268*$c_{2}^{10}$

$G_{2}(c_{2},c_{3})=0=$

23547338655791229204617338928506186026357940595072461145844743562575

17224332611787380814907080082015469486345955534220731803730371551646

24134092641462097641695580678924043455393253397537452244857196895154

46461325491822959772758622053185481177195343628678109510431558799243

57732629782484420615672656492158360322133109432630494581048689547

$-$3097250590883846738286889878288351798304609282451856596410303546337

36319437905844941144694584532179492363486873159383686238249006746894

40160319769886619561748249254067684300497233927857246816557510045452

29549030244795643979218209546091280718239589534262958680509337880753

427578723122948642870383378005580770133313107270509303255691909320*$c_{2}$

$-$1903803609627374035621320978361604020106824484275906499213001947264

95672033886734748658480731859038594412417938078895776898706276221196

35127294306367325389388675757732402447319340359820738619607197620136

52879058280068724250611547838816805478213287696041214145895488077794

7039714261670007939310542108233323980869417122270525843479004885596*$%
c_{2}^{2}$

+8450512390839763624967161124974667624579807298666582130379782638299

901903452364927022770162589839439882861673620370340791082608420840404

311331236303835885178287071688967159857905276468573727498262705288023

535530120799333316114432548548504549670733473680409194666521803651767

1193123383965058588912264378749348068202790932902763246688455431*$c_{2}^{3}$

+4336608402971539347025237252135145455841035812345961936163213074515

668913809930686458789021032092183097107068630380980654723975384806834

399547265746928595076966467194028602153464501847919685450547334048541

838603267464810745721708090718604776111130817877753185112995125825196

9900571656166285598237774354646551889675112530575080545846831937*$c_{2}^{4}$

$-$29667122481091026364901639098151640003965888696689225008930393424968

602525252447427436681754274345497424043473832230915504701237012993339

710812856574652111731271369111860472809751620103390836422947014561918

596742170800703187841128631115509897386033559201960217645649903785096

9895485234230652379590298859552730627182446420246826454883883689*$c_{2}^{5}$

+8713668020397159975202209402993411331481527445489159653742816589285

782666925647284363167235229251382288691898280884602035659684679625961

618445548964513132925721174946043787600439882519732665839121162038859

547938786254023880112832091100880202921766725381340769795219852772659

31253523019517845556699435901635488455864886970958223801767254199*$c_{2}^{6}$

$-$78305675171958244526488263512192141094477411091651712476670327217351

782883871707407108769516955504305407699176179301277874612313054094467

597035177196760703985825236814433985312481352680953166404723733165962

317279471444717834466650359718509004877505793494350037939473327528086

0867972222715638415162988327352914888458003606019978160749679848*$c_{2}^{7}$

$-$50522599203583621625595637115217448838607343066299095645637009117677

845425111468649945267084093503724529301981623817384478696550110017337

119404751862738670370910467517021683484411445974804570961028420156557

202505669968683614281388897157201250167813030135004835015786338672614

037510989474695713072179459238022542102052991592155908191171478*$c_{2}^{8}$

+35315707780022362747687233745930780256946702046940375643458709883815

717091334531092687692414120601353572599659606631610854904141003914876

375870980951760199224617620310327264953286419665073175139662877717976

464011962351866981273603060052461996153251279426523988643704170224549

3505041609179506395307679350757457119876008964575838569723696328*$c_{2}^{9}$

+26196064537923148987259844023868839991472305689605751242386858343654

158387703055190547211883116255260309881480267447955878169277786134938

237906112297170134620139058793181439502365320984021683968720825019581

715365315571505452391632141389861154616280828170929049200565639622657

40795249676124179031191642089155403460321065906554129943039359*$c_{3}$

The polynomials $%
G_{3,6}(s_{6},c_{2})=0,G_{3,5}(s_{5},c_{2})=0,...,G_{3,1}(s_{1},c_{2})=0$
have the same form as $G_{2}(c_{2},c_{3})=0$ given above.

The polynomial $G_{1}(c_{2})=0$ has only four real roots, which are: $-$%
1.871987136, $-0$.3332612900, $-0$.2556946431, 0.2733107997; so the rank of
the dataset is greater than 7. We embed it by joining the following vectors
to the three slices: ${\bf v}_{1}^{\prime }=(1\ \ 0\ \ 0\ \ 0),\ \ {\bf v}%
_{2}={\bf v}_{3}={\bf 0.}$ The embedded dataset is {\bf X}$_{1}^{e}=(${\bf X}%
$_{1}^{\prime }\ \ {\bf v}_{1}^{\prime })^{\prime },$ {\bf X}$_{2}^{e}=($%
{\bf X}$_{2}^{\prime }\ \ {\bf v}_{2}^{\prime })^{\prime }$ \ and \ {\bf X}$%
_{3}^{e}=(${\bf X}$_{3}^{\prime }\ \ {\bf v}_{3}^{\prime })^{\prime }.$ The
rank of the embedded dataset will be calculated by two distinct methods.

First, for the embedded dataset we see that $n\func{col}{\bf \Gamma =}n\func{%
row}{\bf \Gamma =}8,$ so we can calculate the determinant of ${\bf \Gamma }$
as in Example 5, which is:

$det({\bf \Gamma )=}0=$ 111296195967997*$c_{2}^{4}-$163212875913821*$%
c_{2}^{3}-$288078435761246*$c_{2}^{3}$*$c_{3}$

+188384423078426*$c_{2}^{2}$+139757151961919*$c_{2}^{2}$*$c_{3}-$%
123835533958927*$c_{2}^{2}$*$c_{3}^{2}$

+3188520736473*$c_{2}$+1745777654358*$c_{2}$*$c_{3}+$145702375007129*$c_{2}$*%
$c_{3}^{2}$

+154156258186696*$c_{2}$*$c_{3}^{3}-$30068441704134*$c_{3}-$78231890782721*$%
c_{3}^{2}$

$-$9292669314727*$c_{3}^{3}$+24148992371016*$c_{3}^{4}$

Following the argument in Example 5, we note that there is a slight
possibility that there will not be eight distinct values of ($\widetilde{%
c_{2}},\widetilde{c_{3}})$ such that the $det({\bf \Gamma )=}0,$ because it
is of degree 4. So, in general, following this approach of computing we can
not assert that typical rank of generic $7\times 4\times 3$ arrays is $%
\left\{ 7,7+1\right\} .$ However, let us continue our computation as in
Example 5. We obtain the {\bf C} matrix of Lemma 1,rounded to 2 decimal
digits,

{\bf C} := $\left( 
\begin{array}{c}
\lbrack 1,1,1,1,1,1,1,1] \\ 
\lbrack 5.68,63.84,0.6,44.31,3.49,9.33,3.93,21.29] \\ 
\lbrack -38,91,-1,63,-23,-63,-26,30]%
\end{array}%
\right) $

where the third row of {\bf C} represents the randomly generated eight $%
\widetilde{c_{3}}$ values, and the second row represents the corresponding
eight $\widetilde{c_{2}}$ values obtained by solving the $det({\bf \Gamma )=}%
0$ after plugging the $\widetilde{c_{3}}$ values in it.

Now, we can obtain the {\bf S} matrix of Lemma 1 by solving 
\begin{equation}
{\bf s}^{\prime }({\bf X}_{k}^{e}-\widetilde{c_{k}}{\bf X}_{1}^{e})={\bf 0}%
^{\prime }\text{\ for \ }k=2,3  \tag{17}
\end{equation}%
eight times: The $i$th column of {\bf S} corresponds to the eigenvector of
the matrix ${\bf \Gamma (}\widetilde{c_{2i}},\widetilde{c_{3i})}$ associated
with the unique null eigenvalue

{\bf S} := $10^{-3}\times \left( 
\begin{array}{c}
\lbrack -6.85,\ -6.54,\ -4.89,\ -6.51,\ 7.02,\ 6.74,\ 6.97,\ 6.44] \\ 
\lbrack -4.02,\ 5.45,\ 8.94,\ 5.44,\ 3.85,\ 4.11,\ 3.90,\ 5.40] \\ 
\lbrack 9.29,\ -5.99,\ 14.8,\ -5.97,\ -9.49,\ -9.16,\ -9.43,\ -5.91] \\ 
\lbrack -16.4,\ -6.87,\ -34.4,\ -6.83,\ 16.7,\ 16.2,\ 16.6,\ -6.71] \\ 
\lbrack -4.31,\ 0.995,\ -6.31,\ 1.01,\ 4.15,\ 4.40,\ 4.20,\ 1.08] \\ 
\lbrack -11.5,\ -5.37,\ -43.7,\ -5.36,\ 11.8,\ 11.4,\ 11.7,\ -5.32] \\ 
\lbrack -3.22,\ -6.50,\ -6.24,\ -6.50,\ 3.20,\ 3.24,\ 3.20,\ -6.52] \\ 
\lbrack -1000,\ -1000,\ -1000,\ -1000,\ -1000,\ -1000,\ -1000,\ -1000]%
\end{array}%
\right) $

The matrix ${\bf B}^{\prime }={\bf S}^{\prime }{\bf X}_{1}={\bf A}^{-1}{\bf X%
}_{1}$ of Lemma 1 is

{\bf B} := $10^{-2}\times \left( 
\begin{array}{c}
\lbrack 1.06,\ -1.47,\ 22.7,\ -2.11,\ -1.84,\ -0.625,\ -1.61,\ -4.37] \\ 
\lbrack -2.25,\ -1.24,\ -171,\ -1.79,\ 3.73,\ 1.35,\ 3.30,\ -3.72] \\ 
\lbrack 2.49,\ -0.077,\ -22.7,\ -0.111,\ -4.22,\ -1.48,\ -3.70,\ -0.228] \\ 
\lbrack 3.85,\ -0.280,\ -69.4,\ -4.00,\ -6.39,\ -2.31,\ -5.65,\ -0.806]%
\end{array}%
\right) $

And ${\bf A}^{\prime }={\bf S}^{-1};$ finally, we obtain $\widetilde{{\bf X}}%
_{k}={\bf AD(c}_{k}{\bf )B}^{\prime }={\bf X}_{k}.$ This was numerically
verified.

A second approach to compute the matrices {\bf S}, {\bf A}, {\bf B} and {\bf %
C} is via the Gr\"{o}bner basis for the embedded system (17) using the
lexicographic order ($s_{1},s_{2},s_{3},s_{4},s_{5},s_{6},c_{3})$; note that 
$c_{2}$\ is a free variable. The first Gr\"{o}bner basis\ polynomial is

$G_{1}(c_{2},c_{3})=0=$ 111296195967997*$c_{2}^{4}-$163212875913821*$%
c_{2}^{3}-$288078435761246*$c_{2}^{3}$*$c_{3}$

$+$188384423078426*$c_{2}^{2}$+139757151961919*$c_{2}^{2}$*$c_{3}-$%
123835533958927*$c_{2}^{2}$*$c_{3}^{2}$

$+$3188520736473*$c_{2}$+1745777654358*$c_{2}$*$c_{3}+$145702375007129*$%
c_{2} $*$c_{3}^{2}$

$+$154156258186696*$c_{2}$*$c_{3}^{3}-$30068441704134*$c_{3}-$78231890782721*%
$c_{3}^{2}$

$-$9292669314727*$c_{3}^{3}$+24148992371016*$c_{3}^{4},$

which equals $det({\bf \Gamma ).}$ This shows that both approaches are
identical for this particular problem.

\section{\bf Another Application of The Main Lemma}

Consider {\bf nongeneric} dataset of size $4\times 4\times 3$

{\bf X}$_{1}$ := $\left( 
\begin{array}{c}
\lbrack -872410,509152,-155756,301976] \\ 
\lbrack -669515,355308,-105576,215236] \\ 
\lbrack 349983,-898362,265770,-79182] \\ 
\lbrack 3285,-185950,180998,97398]%
\end{array}%
\right) $

{\bf X}$_{2}$ := $\left( 
\begin{array}{c}
\lbrack -403995,481229,24054,201485] \\ 
\lbrack -243133,337616,-4344,94484] \\ 
\lbrack 317091,-174294,-2454,-206076] \\ 
\lbrack -317457,112640,183938,289254]%
\end{array}%
\right) $

{\bf X}$_{3}$ := $\left( 
\begin{array}{c}
\lbrack -274447,214327,-280750,108851] \\ 
\lbrack -252456,116912,-145020,92016] \\ 
\lbrack -127464,-713802,599526,54318] \\ 
\lbrack -38790,-204608,236662,21168]%
\end{array}%
\right) $

To see if the rank of ${\bf X}${\bf \ }is 4, we solve the system (9)
composed of 8 polynomial equations in 5 variables via Gr\"{o}bner basis
using the pure lexicograhic order $(s_{1},s_{2},s_{3},c_{3},c_{2}).$ The
elements of the Gr\"{o}bner basis are

$%
G_{1}(c_{2})=0=-266104+1131869c_{2}+1855673c_{2}^{2}-10091484c_{2}^{3}+3934656c_{2}^{4}; 
$

$%
G_{2}(c_{2},c_{3})=0=70150154675210213-61657878275323159c_{2}-700780737688415568c_{2}^{2} 
$

$+308891236767911424c_{2}^{3}+628616789525725c_{3};$

$G_{3,3}(c_{2},s_{3})=0=-79011958683266608845932181098557$

$%
+37717083374737443006703954200886c_{2}+901077269210427745705210304730192c_{2}^{2} 
$

$%
-390331538460948950190867958454016c_{2}^{3}+1099565644871457602013702982455s_{3}; 
$

$G_{3,2}(c_{2},s_{2})=0=39353103064214280234416428219949$

$%
-190977009897456095062042069799807c_{2}+209064381129999539517569775784236c_{2}^{2} 
$

$%
-57940861139941085694575004742848c_{2}^{3}+5497828224357288010068514912275s_{2}; 
$

$G_{3,1}(c_{2},s_{1})=0=-29281575292540618957256320186316$

$%
-3959967531501755611631756716147c_{2}+390527303469098244882389504161956c_{2}^{2} 
$

$%
-165798278803217428934162052760128c_{2}^{3}+1099565644871457602013702982455s_{1}. 
$

The polynomial $G_{1}(c_{2})=0$ is of degree 4 and it has four real roots,
which are: $-.3369565217,.2929292929,.2962962963$, $2.312500000.$ So the
rank of the dataset is 4 by the main Lemma. Such datasets have been
characterized by their defining equations in Landsberg and Manivel (2006).

\section{INDSCAL arrays}

Let ${\bf X}\in {\bf 
\mathbb{R}
}^{I\times J\times J}$ be a tensor of order 3,where the $i$th slice ${\bf X}%
_{i}\in {\bf 
\mathbb{R}
}^{J\times J}$ for $i=1,...,I$ is symmetric. INDSCAL, proposed by Carroll
and Chang (1970), is a statistical method used in psychometrics to analyse
such arrays. For this reason, we shall name such an array an INDSCAL array
to distinguish it from a general three-way array ${\bf Y}\in {\bf 
\mathbb{R}
}^{I\times J\times K}$ discussed above$,$ where such a decomposition is
usually named PARAFAC, see Harshman (1970). A rank 1 INDSCAL array or a
decomposed tensor is 
\begin{equation}
{\bf D}={\bf a}\otimes {\bf b}\otimes {\bf b},  \tag{18}
\end{equation}%
where ${\bf a}\in {\bf 
\mathbb{R}
}^{I}$\ and ${\bf b}\in {\bf 
\mathbb{R}
}^{J}.$

The following theoretical results are known for generic INDSCAL data ${\bf X}%
\in {\bf 
\mathbb{R}
}^{I\times J\times J}:$ a) By Zellini (1979), see also Rocci and Ten Berge
(1994), if{\bf \ }$I${\bf \ }$\geq ${\bf \ }$J(J+1)/2,\ $then rank$%
(X)=J(J+1)/2.$ b) $I\times 2\times 2$ and $I\times 3\times 3$ arrays are
studied by Ten Berge, Sidiropoulos and Rocci (2004). The rank computation
problem has also been approached from a numerical point of view by Comon and
ten Berge (2008), who applied applied Terracini's lemma, based on the
numerical calculation of the maximal rank of the Jacobian matrix of (2), to
obtain numerically the generic rank of some INDSCAL three-way arrays. The
numerical method based on Terracini's lemma, when used to evaluate rank over 
${\bf 
\mathbb{R}
,}$ gives the generic rank when the typical rank is single-valued, and the
smallest typical rank value otherwise.

For INDSCAL data (7) becomes%
\begin{equation}
{\bf s}^{\prime }{\bf X}_{k}=b_{k}{\bf b}^{\prime }\text{ \ \ for \ }%
k=1,...,J,  \tag{19}
\end{equation}%
for ${\bf X}_{k}\in {\bf 
\mathbb{R}
}^{I\times J},{\bf s}\in {\bf 
\mathbb{R}
}^{I}$ and$\ {\bf b}\in {\bf 
\mathbb{R}
}^{J}$.

We note that (19) has two indeterminacies, scale and sign: It can be
rewritten as $\widetilde{{\bf s}}^{\prime }{\bf X}_{k}=\widetilde{b}_{k}%
\widetilde{{\bf b}}^{\prime }$ \ \ for \ $k=1,...,J,$ , where for instance, $%
\widetilde{{\bf s}}=\lambda {\bf s}$ for any scalar $\lambda >0$ and $%
\widetilde{{\bf b}}=\lambda ^{1/2}{\bf b.}$ The second indeterminacy is the
sign indeterminacy of ${\bf b:}$ replacing ${\bf b}$ by $-{\bf b}$ in (19)
does not change the equality in (19).To eliminate both indeterminacies,
hereafter, we fix 
\begin{equation}
b_{1}=1.  \tag{20}
\end{equation}

We will represent the set of solutions of (19) subject to (20) by $V$
(Veronese variety).

We are interested in the study of the number of solutions of (19) subject to
(20) over ${\bf 
\mathbb{R}
}$ for {\it generic} INDSCAL data for $2\leq J,I\leq J(J+1)/2.$ We
distinguish three cases named, minimal when $I=1+J(J-1)/2$, overdetermined
when $I>1+J(J-1)/2,$ and, underdetermined when $I<1+J(J-1)/2$. The
overdetermined systems is similar to the one discussed above.

{\bf Theorem 3 (minimal system=Veronese variety):} Let $I=1+J(J-1)/2$ and $%
2\leq J\leq I,$ then for generic INDSCAL data the number of roots (real or
complex) of the polynomial system (19) is%
\begin{equation}
degV=2^{J-1}.  \tag{21}
\end{equation}%
{\em Proof: }Let $[b_{1},...,b_{J}]$ be an element of the projective space $%
P^{(J-1)}.$ We note that the right hand side of (19) is a Veronese variety
of degree $d=2$, which is the image of the Veronese map, $\nu _{2},$ defined
by%
\[
\nu _{2}:P^{(J-1)}\rightarrow P^{N}, 
\]%
by sending%
\[
\lbrack b_{1},...,b_{J}]\rightarrow \lbrack
b_{1}^{2},b_{1}b_{2},...,b_{J}b_{J-1},b_{J}^{2}], 
\]%
where the image has $N+1=\left( _{\ \ \ \ 2}^{J-1+2}\right) $ elements
composed of binomials in $b_{1},...,b_{J}.$ While the left hand side of (19)
is a general linear space of projective dimension $I-1.$ The number of
intersections of the general linear space with the Veronese varity is
finite, when $I-1=N-(J-1)$; that is 
\begin{equation}
I=1+J(J-1)/2.  \tag{22}
\end{equation}%
When (22) is true, the finite number of intersections is the degree of the
Veronese variety given in (21), see for instance Harris (1992, p. 231).
\bigskip

{\bf Corollary 1}: The typical rank of INDSCAL arrays with a minimal system
have more than one rank value and the minimum attained value is $I$.

{\em Proof: }For minimal systems and $2\leq J\leq I$, $I\leq degV.$ The rank
of a generic INDSCAL array with a minimal system is $I$, if the number of
real roots of $V$ is greater than or equal to $I$; otherwise its rank is
greater than $I$.

\subsection{Example 7}

We consider a simulated generic dataset of size $4\times 3\times 3$ having
the following 4 slices

{\bf X}$_{1}$ := $\left( 
\begin{array}{c}
\lbrack 54,107,161] \\ 
\lbrack 107,58,13] \\ 
\lbrack 161,13,134]%
\end{array}%
\right) \ ${\bf X}$_{2}$ := $\left( 
\begin{array}{c}
\lbrack 114,-49,-125] \\ 
\lbrack -49,-144,-76] \\ 
\lbrack -125,-76,-8]%
\end{array}%
\right) $

{\bf X}$_{3}$ := $\left( 
\begin{array}{c}
\lbrack -44,7,-48] \\ 
\lbrack 7,-36,-11] \\ 
\lbrack -48,-11,-154]%
\end{array}%
\right) \ ${\bf X}$_{4}$ := $\left( 
\begin{array}{c}
\lbrack 50,92,-4] \\ 
\lbrack 92,100,1] \\ 
\lbrack -4,1,-100]]%
\end{array}%
\right) $

INDSCAL $4\times 3\times 3$ arrays have been studied in detail by Ten Berge,
Sidiropoulos, and Rocci (2004), where it is shown that if a certain
polynomial of degree 4 has 4 real roots, then $rank({\bf X})=4,$ otherwise
the rank is 5.

The Gr\"{o}bner basis with pure lexicographic order given by the following
sequence ($b_{1},b_{2},s_{1},s_{2},s_{3},s_{4})$ of the free variables is
formed of 6 polynomials listed below. The first polynomial $G_{4}(s_{4})=0$
is of degree 4, as shown by ten Berge, Sidiropoulos and Rocci (2004) and
Theorem 3, and it has 2 real roots -0.1881015674e-2, 0.7632125093e-1, so the
rank of the dataset is greater than 4.

$G_{4}(s_{4})=$7337669360341773654444527-4293727819369270858661345768*$s_{4}$

-211863296775796994233864209576*$s_{4}^{2}$-1486920579131214046506874714272*$%
s_{4}^{3}$

+65728692033647334980166673748496*$s_{4}^{4},$

$G_{3}(s_{3},s_{4})=$17560973913802573904674715803175900627366683113948054931

-161223257377160952551452668901669378891965735728005400638*$s_{4}$

-5837696072380594108159410240186154115595431615363079926796*$s_{4}^{2}$

+1318397923701745931624444235931465979525756973974101183796472*$s_{4}^{3}$

+5238806078525191567165441234720094289579153419952152373008*$s_{3}$,

$G_{2}(s_{2},s_{4})=$9628239825303370993360207993471191965769478430007385299

+11068252823433558277754489464864186098568340630108319931766*$s_{4}$

+705995008022931051200292932727627624011448550604188031890116*$s_{4}^{2}$

-9118208129962992736609222153263187238797217283937992554624760*$s_{4}^{3}$

+20955224314100766268661764938880377158316613679808609492032*$s_{2}$,

$G_{1}(s_{1},s_{4})=$-9384923940854434492010502183000235854601288296806481877

-1964822700901995622589398714750515451903512627004147123640*$s_{4}$

+1914122466041979887104614509203684754556718777770379683036*$s_{4}^{2}$

+689540537276652567152783462787012635706612460537819392520176*$s_{4}^{3}$

+2619403039262595783582720617360047144789576709976076186504*$s_{1}$,

$G_{2,4}(b_{2},s_{4})=$%
-155028823048701914654384720617223421617571775035134363431

-66211886029016478638439782073183667186014332650277499221046*$s_{4}$

-3324980827880602990137773057985895339331874506372137213739628*$s_{4}^{2}$

+44167907819442655873419676087304190072683931413772512430408456*$s_{4}^{3}$

+1309701519631297891791360308680023572394788354988038093252*$b_{2}$,

$G_{1}(b_{1},s_{4})=$%
-1514819909584866108143372567736179608809277026174154396973

-345404316274377016240902795956549557726567542396897273802586*$s_{4}$

-7696649874609748839698115121543942313397062924509594873506300*$s_{4}^{2}$

+161342893355178852699731325084429928324620333696539365418650824*$s_{4}^{3}$

+1905020392190978751696524085352761559846964879982600862912*$b_{1}$

\subsection{Example 8}

We consider a simulated generic INDSCAL dataset of size $7\times 4\times 4$
having the following $7$ slices

{\bf X}$_{1}$ := $\left( 
\begin{array}{c}
\lbrack 140,86,-110,-4] \\ 
\lbrack 86,-182,70,36] \\ 
\lbrack -110,70,104,183] \\ 
\lbrack -4,36,183,148]%
\end{array}%
\right) \ ${\bf X}$_{2}$ := $\left( 
\begin{array}{c}
\lbrack -20,100,173,-56] \\ 
\lbrack 100,128,101,75] \\ 
\lbrack 173,101,124,65] \\ 
\lbrack -56,75,65,-158]%
\end{array}%
\right) $

{\bf X}$_{3}$ := $\left( 
\begin{array}{c}
\lbrack 178,15,-186,52] \\ 
\lbrack 15,196,119,-148] \\ 
\lbrack -186,119,-138,43] \\ 
\lbrack 52,-148,43,-110]%
\end{array}%
\right) \ ${\bf X}$_{4}$ := $\left( 
\begin{array}{c}
\lbrack -8,-137,21,20] \\ 
\lbrack -137,-60,64,5] \\ 
\lbrack 21,64,-24,-14] \\ 
\lbrack 20,5,-14,-128]%
\end{array}%
\right) $

{\bf X}$_{5}$ := $\left( 
\begin{array}{c}
\lbrack 194,-164,-36,-6] \\ 
\lbrack -164,2,-114,-110] \\ 
\lbrack -36,-114,64,-127] \\ 
\lbrack -6,-110,-127,-18]%
\end{array}%
\right) \ ${\bf X}$_{6}$ := $\left( 
\begin{array}{c}
\lbrack -22,74,85,-40] \\ 
\lbrack 74,-198,23,-53] \\ 
\lbrack 85,23,-152,18] \\ 
\lbrack -40,-53,18,-48]%
\end{array}%
\right) $

{\bf X}$_{7}$ := $\left( 
\begin{array}{c}
\lbrack -94,109,-16,90] \\ 
\lbrack 109,124,164,-93] \\ 
\lbrack -16,164,98,-134] \\ 
\lbrack 90,-93,-134,184]%
\end{array}%
\right) $

The Gr\"{o}bner basis with pure lexicographic order given by the following
sequence ($s_{1},...,s_{7},b_{1},b_{2},b_{3})$ of the free variables is
formed of 10 polynomials, but only the first one is shown below. The first
polynomial $G_{3}(b_{3})=0$ is of degree 8, as shown in Theorem 3, and it
has 2 real roots -4.615952848, 1.035693119, so the rank of the dataset is
greater than 7.

$G_{3}(b_{3})=$%
-267319790697212354162205439965563724346086890209668628287611296

-418573483979735109514695930195818332286961955303805928337210144*$b_{3}$

-53224562968122644847846329140305933491773608156555442814188832*$b_{3}^{2}$

+190260260230311283947025128614232688395842607775283676963072480*$b_{3}^{3}$

+172806709139583658797792038234309205181892588602072553465939944*$b_{3}^{4}$

+164461253658569745584828839860332360925297521683057843681458288*$b_{3}^{5}$

+95090716874891491062104108972298040778579595301835996945880112*$b_{3}^{6}$

+24575461542028106507015735163996805821952192308876843008456228*$b_{3}^{7}$

+2240887382441309183839416634048576470976843441637962999441259*$b_{3}^{8}$

\section{Conclusion}

We introduced a new method to compute ranks of three-way arrays, by showing
that it is intimately related to the solution set of a system of polynomial
equations, which is a well developed and active area of mathematics known as
algebraic geometry. The new method was used to compute numerically the ranks
of some sizes of three-way arrays over ${\bf 
\mathbb{R}
}$ via Gr\"{o}bner basis.

The problem of computing the rank of overdetermined systems by solving
embedded polynomial systems is a work in progress{\bf .}

\medskip \medskip

{\bf Acknowledgements}

This research was financed by the Natural Sciences and Engineering Research
Council of Canada. \ The author thanks Pr. J. Ten Berge and an unknown
reviewer for their constructive comments, and Marc Sormany for help in
programming.\medskip

{\bf Appendix 1}

Below the matrix ${\bf Y}k={\bf X}_{k}^{\prime }.$

\TEXTsymbol{>} K := 3;

\TEXTsymbol{>} J := 8;

\TEXTsymbol{>} I := 15;

\TEXTsymbol{>} with(LinearAlgebra);

\TEXTsymbol{>} Y1 := RandomMatrix(J, I);

\TEXTsymbol{>} Y2 := RandomMatrix(J, I);

\TEXTsymbol{>} Y3 := RandomMatrix(J, I);

\TEXTsymbol{>} S := Vector(1 .. I, 1);

\TEXTsymbol{>} for h from 1 to I-1 do

\ \ S[h] := s$_{h}$ end do;

\TEXTsymbol{>} M1 := Y2-c$_{2}$*Y1;

\TEXTsymbol{>} P1 := M1.S;

\TEXTsymbol{>} M2 := Y3-c$_{3}$*Y1;

\TEXTsymbol{>} P2 := M2.S;

\TEXTsymbol{>} poly := [seq(P1[l], l = 1 .. J), seq(P2[n], n = 1 .. J)];

\TEXTsymbol{>} with(Groebner);

\TEXTsymbol{>} liste := seq(s$_{h}$, h = 1 .. I-1);

\TEXTsymbol{>} polyG := Basis(poly, plex(liste, c$_{3}$, c$_{2}$));

\TEXTsymbol{>} polyG[1];

\TEXTsymbol{>} S := [solve(polyG[1])];

\TEXTsymbol{>} nops(S);

\TEXTsymbol{>} fsolve(polyG[1]);

\TEXTsymbol{>} nops([fsolve(polyG[1])]);

\TEXTsymbol{>} fsolve(polyG[1], a, complex);

\TEXTsymbol{>} nops([fsolve(polyG[1], c$_{3}$, complex)]);\bigskip

{\bf References}

Abo, H., Ottaviani, G., Peterson, C. (2006). Induction for secant varieties
of Segre varieties. arXiv:math.AG/0607191 v3, 2 August.

Basu, S., Pollack, R. and Roy, M-F. (2006). Algorithms in Real Algebraic
Geometry. Springer, Berlin, 2nd edition.

B\"{u}rgisser, P., Clausen, M., Shokrollahi, M.A. (1997). Algebraic
Complexity Theory, with the collaboration of Thomas Lickteig. {\it %
Grundlehren der Mathematischen Wissenschaften}, 315. Springer-Verlag, Berlin.

Carroll, J.D. and Chang J.J. (1970). Analysis of individual differences in
multidimensional scaling via an n-way generalization of Eckart-Young
decomposition. {\it Psychometrika}, 35, 283-319.

Catalisano, M.V., Geramita, A.V., Gimigliano, A. (2002). Ranks of tensors,
secant varieties and fat points. {\it Linear Algebra Appli}. 355, 263-285.
Erratum, {\it Linear Algebra Appli}. 367 (2003), 347-348.

Comon, P. and ten Berge, J. (2008). Generic and typical ranks of three-way
arrays. arXiv:0802.2371v1[cs.OH], 17 February.

Cox, D., Little, J. and O'Shea, D. (2007). Ideals, Varities, and Algorithms.
Third edition, Springer, N.Y.

Friedland, S. (2008). On the generic rank of 3-tensors.
ArXiv:0805.3777v2[math.AG], 27 May.

Harris, J. (1992). {\it Algebraic Geometry: A first course. }Springer, N.Y%
{\it .}

Harshman, R. (1970). Foundations of the PARAFAC procedure: Models and
conditions for an \textquotedblleft explanatory\textquotedblleft\ multimodal
factor analysis. {\it UCLA Working Papers in Phonetics}, 16, 1-84.

Ja' Ja', J. (1979). Optimal evaluation of pairs of bilinear forms. {\it SIAM
Journal of Computing}, 8, 443-462.

Kruskal, J.B. (1977). Three-way arrays: Rank and uniqueness of trilinear
decompositions with applications to arithmetic complexity and statistics. 
{\it Linear Algebra and its Applications}, 18, 95-138.

Kruskal, J.B. (1983). Statement of some current results about three-way
arrays. Unpublished manuscript, AT\&T Bell Laboratories, Murray Hill, N.J,
downloadable from
http://www.leidenuniv.nl/fsw/three-mode/biblogr/biblio\_k.htm

Kruskal, J.B. (1989). Rank, decompostion, and uniqueness for 3-way and N-way
arrays. In R. Coppi \& S. Bolasco (Eds.), {\it Multiway data Analysis} (pp.
7-18). Amsterdam: North Holland.

Landsberg, J.M. and Manivel, L. (2006). Generalizations of Strassen's
equations for secant varieties of Segre varieties.
ArXiv:math/0601097v1[math.AG], 5 January.

Martin,M C. (2004). Tensor Decompositions Workshop Discussion Notes,
American Institute of Mathematics (AIM), available online.

Rocci, R. and Ten Berge, J. (1994). A simplification of a result by Zellini
on the maximal rank of symmetric three-way arrays. {\it Psychometrika}, 59,
377-380.

Strassen, V. (1983). Rank and optimal computations of generic tensors. {\it %
Linear Algebra Appl}. 52/53, 645-685.

Sturmfels, B. (2002). {\it Solving Systems of Polynomial Equations}. CBMS
Reg. Conf. Ser. in Mathematics, 97, AMS.

Ten Berge, J.M.F. (1991). Kruskal's polynomial for $2\times 2\times 2$
arrays and a generalization to $2\times n\times n$ arrays. {\it Psychometrika%
}, 56, 631-636.

Ten Berge, J.M.F. (2000). The typical rank of tall three-way arrays. {\it %
Psychometrika}, 65, 525-532.

Ten Berge, J.M.F. (2004a). Partial uniqueness in CANDECOMP\TEXTsymbol{%
\backslash}PARAFAC. {\it Journal of Chemometrics}, 18, 12-16.

Ten Berge, J.M.F. (2004b). Simplicity and typical rank of three-way arrays,
with applications to Tucker-3 analysis with simple cores. {\it Journal of
Chemometrics}, 18, 17-21.

Ten Berge, J.M.F. and Kiers, H.A.L. (1999). Simplicity of core arrays in
three-way principal component analysis and the typical rank of $P\times
Q\times 2$ arrays. {\it Linear Algebra its Applications}, 294, 169-179.

Ten Berge, J.M.F. and Stegeman, A. (2006). Symmetry transformations for
square sliced three-way arrays, with applications to their typical rank. 
{\it Linear Algebra its Applications}, 418 (1), 215-224.

Ten Berge, J.M.F., Sidiropoulos, N.D. and Rocci, R. (2004). Typical rank and
INDSCAL dimensionality for symmetric three-way arrays of order $I\times
2\times 2$ or $I\times 3\times 3$. {\it Linear Algebra its Applications},
388, 363-377.

Zellini, P. (1979). On the optimal computation of a set of symmetric and
persymmetric bilinear forms. {\it Linear Algebra its Applications}, 23,
101-119.

/end

\end{document}